\documentclass[12pt,preprint]{aastex}
\begin{document}

\title{Double-Damped Lyman Alpha Absorption: A Possible
Large Neutral Hydrogen Gas Filament Near Redshift $z=1$\altaffilmark{1}}

\author{David A. Turnshek\altaffilmark{2}, Sandhya M. Rao\altaffilmark{2}, 
Daniel B. Nestor\altaffilmark{2}, \\ Daniel Vanden Berk\altaffilmark{2}, 
Mich\`ele Belfort\altaffilmark{2}, and Eric M. Monier\altaffilmark{3}}

\altaffiltext{1}{Based on data obtained from the Sloan Digital
Sky Survey (SDSS) and on observations made with the Hubble Space
Telescope (HST) operated by STScI-AURA for NASA, the MMT (using 
NOAO public access time) operated by the Smithsonian Institution 
and the University of Arizona, and the Infrared Telescope Facility 
(IRTF) operated by the University of Hawaii IfA for NASA.}

\altaffiltext{2}{Dept. of Physics \& Astronomy, University of 
Pittsburgh, Pittsburgh, PA 15260
(turnshek@pitt.edu, rao@everest.phyast.pitt.edu,
dbn@phyast.pitt.edu, danvb@phyast.pitt.edu, mbelfort@phyast.pitt.edu)}

\altaffiltext{3}{Dept. of Astronomy, Ohio State University, Columbus, OH 43210 
(monier@astronomy.ohio-state.edu)}

\begin{abstract}

We report the discovery of two damped Ly$\alpha$ absorption-line systems
(DLAs) near redshift $z=1$ along a single quasar sightline (Q1727+5302)
with neutral hydrogen column densities of $N_{HI} =
(1.45\pm0.15)\times10^{21}$ and $(2.60\pm0.20)\times10^{21}$
atoms cm$^{-2}$.  Their sightline velocity difference of 
13,000 km s$^{-1}$ corresponds to a proper separation
of 106$h_{70}^{-1}$ Mpc if interpreted as the Hubble flow
($\Omega_m=0.3$, $\Omega_\Lambda=0.7$).  
The random probability of such an occurrence
is significantly less than 3\%. Follow-up spectroscopy reveals 
neutral gas-phase Zn abundances of [Zn/H] = $-0.58\pm0.15$
(26.5\% solar) and $-1.32\pm0.28$ (4.7\% solar), respectively.
The corresponding Cr abundances are [Cr/H] = $-1.26\pm0.15$ (5.5\%
solar) and $-1.77\pm0.28$ (1.7\% solar), respectively, which is
evidence for depletion onto grains.  Follow-up IR images
show the two most likely DLA galaxy candidates
to have impact parameters of $\approx22h_{70}^{-1}$
kpc and $\approx32h_{70}^{-1}$ kpc if near $z=1$. They are 
significantly underluminous relative to the galaxy population at
$z=1$.  To investigate the possibility of additional high-$N_{HI}$
absorbers we have searched the SDSS database for $z>1$
quasars within 30 arcmin of the original sightline.  Five were found,
and two show strong \ion{Mg}{2}-\ion{Fe}{2} absorption near
$z=1$, consistent with classical DLA absorption $\approx 37$\% of 
the time, but almost always $N_{HI} > 10^{19}$ atoms cm$^{-2}$.
Consequently, this rare configuration of four high-$N_{HI}$ absorbers
with a total sightline velocity extent of 30,600 km s$^{-1}$ may
represent a large filament-like structure stretching over a proper
distance of 241$h_{70}^{-1}$ Mpc along our sightline, and a region
in space capable of harboring excessive amounts of neutral gas. 
Future studies of this region of the sky are encouraged.

\end{abstract}

\keywords{galaxies: clusters --- large-scale 
structure of universe --- quasars: absorption lines --- quasars:
individual (Q1725+5254, Q1727+5302, Q1727+5311)}

\section{Introduction}

Intervening damped Ly$\alpha$ absorption-line systems (DLAs) in
quasar spectra are very rare, with an incidence of $\approx 0.17$
per unit redshift at $z=1$ (Rao, Turnshek, \& Nestor 2004, hereafter
RTN2004). Consequently, unless DLAs are correlated, the appearance
of two DLAs along any single quasar sightline (``double-damped'')
represents a very unlikely event. As such, the discovery of any
double-damped absorption  warrants a closer investigation.

Here we report the discovery of double-damped absorption near $z=1$
in the Sloan Digital Sky Survey (SDSS) quasar Q1727+5302 during our
most recent {\it Hubble Space Telescope} UV spectroscopic survey for
DLAs (RTN2004).  The purpose of the present paper is to report initial
results pertaining to this discovery, and thereby encourage future
studies of this region of the sky.  The velocity separation of the
absorption is 13,000 km s$^{-1}$, which corresponds to a proper
radial distance  of $106h^{-1}_{70}$ Mpc if interpreted as due
to the Hubble flow.\footnote{To calculate physical quantities in 
this paper we adopt a cosmology with $\Omega_m=0.3$,
$\Omega_\Lambda=0.7$, and $H_0=70$ km s$^{-1}$ Mpc$^{-1}$ ($h_{70}
= H_0/70$).} We speculate that this configuration may represent
a neutral hydrogen gas filament with a large cosmological extent
along our sightline. In fact, the comoving size of this putative
filament would be larger than anything previously reported.

As discussed in earlier contributions (e.g., Rao \& Turnshek 2000,
hereafter RT2000, and references therein), DLAs are excellent
tracers of the bulk of the neutral hydrogen gas in the Universe,
and the aim of our most recent DLA UV survey has been to improve our
knowledge of the incidence and cosmological mass density of DLAs at
redshifts $z<1.65$.  The new sample which led to the discovery of the
double-damped absorption was derived from the 
SDSS Early Data Release (EDR) (Schneider et al. 2002).
We applied a strong \ion{Mg}{2}-\ion{Fe}{2} rest equivalent width
(REW) selection criterion (RT2000) to optical spectra in order to
identify candidate DLA absorption lines ($N_{HI} \ge 2\times10^{20}$
atoms cm$^{-2}$), and then we obtained HST STIS UV spectra to
confirm or refute their presence.  The current overall success rate
for identifying DLAs with this method is $\approx 37$\%.  Since the
\ion{Mg}{2}$\lambda\lambda$2796,2803 absorption lines are saturated,
the REW of the absorption is most closely tied to kinematic spread,
not column density.  Recently, Nestor et al. (2003) have discussed
evidence for a correlation between kinematic spread and metallicity.

The paper is organized as follows.  In \S2 we present the HST
discovery spectrum for the double-damped absorption, a follow-up
MMT spectrum used to determine neutral-gas-phase metal abundances,
and IRTF imaging data used to search for
galaxies associated with the double-damped
absorbers along the quasar sightline. In \S3 we summarize evidence 
that exists for strong \ion{Mg}{2} absorption systems near $z=1$ in 
other SDSS quasars in the same region of the sky. A brief summary and 
discussion of the results is presented in \S4.

\section{Observations}

Q1727+5302 (SDSS J172739.03+530229.16) is the J2000 coordinate
designation of the quasar which exhibits double-damped
absorption near $z=1$.  The quasar has an emission redshift of
$z_{em}=1.44$ and an SDSS g-band magnitude of 18.3.  A search for
the \ion{Mg}{2}$\lambda\lambda$2796,2803 absorption doublet in
bright ($g<19$) SDSS EDR quasars  resulted in the identification
of two systems along this sightline at $z=0.9448$ and $z=1.0312$.
Table 1 gives REWs of some of the identified metal absorption
lines.

\subsection{Spectroscopy: \ion{H}{1} Column Densities and Element Abundances}

The spectrum shown in Figure 1 was obtained during a 73 minute
exposure with the HST on 1 January 2003 in the STIS NUV G230L
mode.  See RTN2004 for details of the observing program.
The DLA lines at $z=0.9465$ and $z=1.0315$ are visible.
The insert to the figure shows two Voigt damping profiles
that have been fitted simultaneously to the two Ly$\alpha$ lines.
These have neutral hydrogen column densities of $N_{HI} = (1.45
\pm 0.15) \times10^{21}$ atoms cm$^{-2}$ and $N_{HI} = (2.60 \pm
0.20) \times10^{21}$ atoms cm$^{-2}$, respectively. The $1\sigma$
errors were determined by assessing uncertainties in the continuum
fit as described in RT2000.

We used the 6.5-m MMT on 2 July 2003 to obtain spectroscopic
observations of both systems making up the double-damped absorption
in order to determine their neutral gas phase metal abundances. We
used the \ion{Zn}{2} and \ion{Cr}{2} lines (Figure 2),
as well as \ion{Si}{2}, \ion{Fe}{2}, and \ion{Mn}{2} lines
(not shown).  The  method employed to measure the metal abundances is
the same as that used by Nestor et al. (2003).  Relative to the solar
measurements of Grevesse \& Sauval (1998), we find Zn abundances of
[Zn/H] = $-0.58 \pm 0.15$ (26.5\% solar) and $-1.32 \pm 0.28$ (4.7\%
solar) for the $z=0.9448$ and $z=1.0312$ systems, respectively.
The corresponding Cr abundances are  [Cr/H] = $-1.26 \pm 0.15$
(5.5\% solar) and $-1.77 \pm 0.28$ (1.7\% solar), so there is
evidence for depletion onto grains. These determinations fall within
the range of DLA metallicities that have been reported near $z=1$
(e.g., see Prochaska et al. 2003). Also,
for the $z=0.9448$ system we find
[Si/H] = $-0.81 \pm 0.13$ (15.6\% solar),
[Fe/H] = $-1.37 \pm 0.093$ (4.3\% solar)\footnote{To calculate Fe abundances 
we used the FeII f-values in the NIST database reported at 
physics.nist.gov/cgi-bin/AtData/display.ksh in March 2004, whereas for 
the other elements we used those adopted by Nestor et al. (2003) which 
were compiled at kingpin.ucsd.edu/$\sim$hiresdla by Jason Prochaska.}, 
and [Mn/H] = $-1.18 \pm 0.076$ (6.6\% solar); for the $z=1.0312$ system we find 
[Si/H] = $-1.46 \pm 0.14$ (3.5\% solar), [Fe/H] = $-2.10 \pm 0.13$ 
(0.8\% solar)$^2$, and [Mn/H] = $-1.89 \pm 0.15$ (1.3\% solar). 

\subsection{Infrared Imaging: Luminous Objects Along the Sightline}

Some of our recent imaging results on low-redshift DLAs have
been reported by Rao et al. (2003).  On 2 April 2003 we made similar
IR (JHK bands) observations of the double-damped
sightline with the 3.0-m IRTF.  Figure 3 shows a 14\arcsec\ x 14\arcsec\
section of the H-band image centered on the quasar.
The limiting 1$\sigma$ surface brightness is 22.0 H
magnitudes per square arcsec.  Objects labeled G1 and G2 are reasonable
candidates for the DLA galaxies based on their proximity to
the sightline.  If at $z=1$, their impact parameters of
$\approx 2.7\arcsec$ and $\approx 4.0\arcsec$ correspond to proper
transverse distances of $\approx 22h^{-1}_{70}$ kpc and $\approx
32h^{-1}_{70}$ kpc, respectively.  G1 and G2 have H-band magnitudes
of $m_H(\rm G1)=21.1\pm0.3$ and $m_H(\rm G2)=20.7\pm0.2$. They are
also visible in our J-band and K-band images (not shown), with
magnitudes $m_J(\rm G1)=21.5\pm0.2$, $m_J(\rm G2)=21.7\pm0.2$,
$m_K(\rm G1)=21.0\pm0.4$, and $m_K(\rm G2)=20.0\pm0.2$.

Adopting $L_K^*=-24.1$ from local galaxies studies (e.g., Bell et
al. 2003), a $k$-correction appropriate for an Sa type galaxy from
Poggianti (1997), and assuming no evolution, a local $L_K^*$ galaxy
redshifted to $z=1$ would have $m_K=19.5$ for our adopted cosmology.
We note that at $z=1$ $k$-corrections for all galaxy types are
relatively small in the K-band, and the magnitudes of G1 and G2
would correspond to luminosities of $\approx0.25L_K^*(local)$ and
$\approx 0.6L_K^*(local)$, respectively, if they were at $z=1$.
However, recently Ellis \& Jones (2004) have presented K-band
observations of the galaxy population residing in three x-ray
selected massive clusters at $z=0.8-1.0$.  Fits to their data
indicate that at redshifts near $z=1$ the galaxy population
in massive clusters has $K^*$ lying in the range $17.6 - 18.4$
mag. Thus, $L_K^*(z=1)$ is $\approx 1.5\pm0.4$ mag brighter than
$L_K^*(local)$, which is what might be expected from passive
evolution.  Therefore, G1 and G2 are $\approx0.06L_K^*(z=1)$ and
$\approx0.15L_K^*(z=1)$, respectively, and it should be emphasized
that there are no good candidate DLA galaxies with luminosities
near $L_K^*(z=1)$. Consequently, if G1 and G2 are passively evolving
DLA galaxies, they can simply be interpreted as the progenitors of
local dwarfs. 

We note that if G1 was at the quasar redshift 
($z_{em}=1.44$), it would be 
$\approx0.6L_K^*(local)$.  However, G1 is offset from the quasar, so
if it were related to the quasar it would likely be associated
with an interaction. Low-redshift quasar host galaxies
are generally several times $L^*(local)$ (e.g., Hamilton, Casertano,
\& Turnshek 2002; Jahnke \& Wisotzki 2003).

\section{Search for Strong \ion{Mg}{2} in Other SDSS Quasars within 30 arcmin} 

One way to test the hypothesis that the double-damped absorption
represents an extended filament along our sightline  is to search
for absorption near $z=1$ in other quasars
with $z_{em}>1$ that lie approximately in the same direction.
This possibility can be considered since the
SDSS has searched this region of the sky for quasars.  Five other
$z_{em}>1$ quasars are known that meet the criteria of having spectra
of high enough quality to detect strong \ion{Mg}{2}-\ion{Fe}{2}
systems and lying within 30 arcmin of the double-damped sightline.
An angular size of 30 arcmin (SDSS data presently cut off beyond
this on one edge) is considerably smaller than the putative
$106h_{70}^{-1}$ Mpc (proper distance) long filament at $z=1$,
but this is reasonable if we are viewing cosmic structure along a
filament. All five of these other quasars fall below the brightness
limits used for inclusion in our current HST
UV survey for low-redshift DLAs (RTN2004). Three of the quasars
(SDSS Q1727+5301 at z$_{em}=1.60$, SDSS Q1727+5306 at z$_{em}=1.97$,
and SDSS Q1729+5312 at z$_{em}=1.33$) do not show evidence for strong
\ion{Mg}{2} absorption near $z=1$, so there is little chance that
they have DLA absorption near $z=1$.  However, the remaining two do
show strong \ion{Mg}{2}-\ion{Fe}{2} absorption which meet the criterion 
for selecting DLA candidates (RT2000). The sightline locations and
presence of absorption may constrain
the geometry of any filament. The
two new absorption systems (REWs are reported in Table 1) are at $z=0.9706$ (in SDSS Q1725+5254 at
$z_{em}=1.36$) and $z=1.1536$ (in SDSS Q1727+5311 at $z_{em}=1.80$),
separated from the double-damped sightline by $\approx 25$ arcmin
and $\approx 10$ arcmin, respectively. Thus, there are four strong
\ion{Mg}{2}-\ion{Fe}{2} absorption systems that lie along similar
sightlines, within a redshift spread of $\Delta z = 0.2088$,
corresponding to a velocity separation of $30,600$ km s$^{-1}$.
Given our strong \ion{Mg}{2}-\ion{Fe}{2}
selection criterion, if these new systems are not DLAs, they are
almost certainly sub-DLAs with \ion{H}{1} column densities $>
10^{19}$ atoms cm$^{-2}$.  Figure 4 shows continuum-normalized
regions of the SDSS spectra for all three quasars with strong
\ion{Mg}{2}-\ion{Fe}{2} absorption.

\section{Summary and Discussion}

Investigations of the possibility of clustering between DLAs
and Lyman break galaxies (LBG) have been made at high redshift
(e.g., Gawiser et al. 2001, Aldelberger et al. 2003, Bouch\'e \&
Lowenthal 2003).  Gawiser et al. (2001) and Aldelberger et al. (2003)
find no significant evidence for clustering at $z\approx4$ and
$z\approx3.2$, respectively, while Bouch\'e \& Lowenthal (2003)
present weak evidence ($2.6\sigma$ significance) for clustering
at $z\approx3$ on a size scale up to $\approx1.8h^{-1}_{70}$ Mpc
($\approx7h^{-1}_{70}$ Mpc comoving).

Here we have reported the discovery of an apparently non-random
(see below) structure on the sky near redshift $z=1$ (\S2.1) which
is much larger than the scales explored in high-redshift DLA-LBG
clustering studies and much larger than normal galaxy clustering.
The structure consists of two extremely high-$N_{HI}$ (even by DLA
standards) DLA absorbers separated by 13,000 km s$^{-1}$ along a
single quasar sightline. If interpreted as a cosmological redshift
separation, this double-damped structure has a radial proper
distance of $106h_{70}^{-1}$ Mpc. The incidence of DLAs above a
survey threshold of $N_{HI} \ge 2\times10^{20}$ atoms cm$^{-2}$
is $\approx 0.17$ per unit redshift at $z=1$ (RTN2004).  Given this
incidence, the probability of observing a second DLA within $\Delta z
\approx \pm0.09$ of another DLA is $<$3\%. We quote this probability
as an upper limit because the incidence of DLAs is significantly
smaller for higher column density systems, and the column densities
of the two systems included in the double-damped absorption are
factors of $\approx7$ and $\approx13$ times larger than the survey
threshold.$\footnote{Of the $\approx50$ DLAs presently
known to us at $z<1.65$, the two which make up
the double-damped absorption rank as the 8th and 14th
highest $N_{HI}$ systems.}$ Therefore, the double-damped
absorption may be the result of correlated DLAs and be caused by a
cosmologically extended filament of neutral gas along our sightline.

In addition to the identification of this remarkable structure,
we have made metal abundance determinations for the two systems
which make up the double-damped absorption. We find them to have [Zn/H]
= $-0.58\pm0.15$ (26.5\% solar) and $-1.29\pm0.27$ (4.7\% solar),
with evidence for some depletion onto grains in both cases (\S2.1).
Infrared imaging indicates that the two most likely DLA candidate
galaxies are relatively faint in relation to the galaxy population at
$z=1$, with $K$-band luminosities that are $\approx0.06L_K^*(z=1)$
and $\approx0.15L_K^*(z=1)$ (\S2.2).  These add to the list of
underluminous galaxies that have been identified as being responsible
for DLA absorption (Rao et al. 2003). The results indicate that the 
presence of luminous galaxies relative to the local population evidently
is not a requirement for the presence of large concentrations of neutral
gas. We have also identified
two new candidate DLA systems in this same region of the sky,
separated from the original sightline by 10 arcmin and 25 arcmin
(\S3). The discovery of these two additional systems increases the
probability that this is a non-random structure.  If the two new
redshift systems are included, the structure stretches $30,600$
km s$^{-1}$ along the sightline, corresponding to a radial proper
distance of $241h_{70}^{-1}$ Mpc. 

Filaments nearly as large as that implied by the double-damped absorption
have been seen in mock redshift surveys
and cold dark matter simulations of structure formation. For
example, Faltenbacher et al. (2002) report correlations in cluster
orientations with respect to one another and find 
alignments of galaxy clusters' major axes on comoving scales of
$\approx 140h_{70}^{-1}$ Mpc, corresponding to a proper distance
of $\approx 70h_{70}^{-1}$ Mpc at $z=1$.  The large scale structure
simulations of Eisenstein, Loeb, \& Turner (1997) show similar
alignments.

Recently, Palunas et al. (2004) have reported evidence for
a structure with a proper size of $\approx25h^{-1}_{70}$ Mpc
($\approx80h^{-1}_{70}$ Mpc comoving), which they found during a
search for Ly$\alpha$-emitting galaxies at $z=2.38$.  Miller et
al. (2004) have reported evidence for structures on even larger
scales based on an analysis of the QSO distribution in the 2dF
redshift survey.  For this case of double-damped absorption, the 
size of the putative neutral hydrogen gas filament would be larger 
than any claimed so far.  Therefore, our interpretation should be considered 
speculative pending future studies. Nevertheless, the properties of
the double-damped absorption are of course relevant to 
studies of DLAs in general. 

The present-day cosmological model (approximately 73\% dark energy,
24\% dark matter and 5\% ordinary matter, e.g., Spergel et al. 2003)
is one in which large-scale structures can form early in time.
However, a specific set of cosmological parameters may also
indicate that it is highly improbable for certain structures to
grow from initial Gaussian perturbations.  Thus, surveys to find
evidence for extreme large-scale structure at high redshift have
the potential to result in important cosmological constraints.
Based on the numbers of DLAs discovered at low redshift so far and 
the DLA column density distribution, the existence of this 
double-damped absorption along one sightline represents evidence 
for a non-random distribution of DLAs which should be further investigated.

\bigskip

\centerline{\bf Acknowledgments}

We thank members of the SDSS collaboration who made the SDSS project
a success and who made the EDR spectra available. 
We acknowledge support from NASA-STScI, NASA-LTSA,
and NSF. HST-UV spectroscopy made the $N_{HI}$ determinations
possible, while follow-up metal abundance measurements (MMT) and
imaging (IRTF) were among the aims of our LTSA and NSF programs.
Funding for creation and distribution of the SDSS Archive has
been provided by the Alfred P. Sloan Foundation, Participating
Institutions, NASA, NSF, DOE, the Japanese Monbukagakusho, and the
Max Planck Society. The SDSS Web site is www.sdss.org. The
SDSS is managed by the Astrophysical Research Consortium for
the Participating Institutions: University of Chicago, Fermilab,
Institute for Advanced Study, the Japan Participation Group,
Johns Hopkins University, Los Alamos National Laboratory, the
Max-Planck-Institute for Astronomy (MPIA), the Max-Planck-Institute
for Astrophysics (MPA), New Mexico State University, University
of Pittsburgh, Princeton University, the United States Naval
Observatory, and University of Washington.

%\newpage

\clearpage

\begin{center}
\begin{deluxetable}{lcccccc}
\tablewidth{0pc}
%\doublespace
\tablenum{1}
\tablecaption{Metal Absorption Line Rest Equivalent Width Measurements}
\tablehead{
\colhead{Quasar} &
\colhead{$z_{\rm em}$} &
\colhead{$z_{\rm abs}$} &
\colhead{$W^{\lambda2600}_{0}$ (\AA)} &
\colhead{$W^{\lambda2796}_{0}$ (\AA)} &
\colhead{$W^{\lambda2803}_{0}$ (\AA)} &
\colhead{$W^{\lambda2852}_{0}$ (\AA)} \\[.2ex]
\colhead{} &
\colhead{} &
\colhead{} &
\colhead{\ion{Fe}{2}} &
\colhead{\ion{Mg}{2}} &
\colhead{\ion{Mg}{2}} &
\colhead{\ion{Mg}{1}}
}
\startdata
Q1727+5302 & 1.44   & 0.9448 & 2.19(0.13) & 2.83(0.07) & 2.51(0.07) & 0.99(0.07) \nl 
\nodata    & \nodata & 1.0312 & 0.76(0.11) & 0.92(0.06) & 1.18(0.08) & 0.33(0.10) \nl
Q1725+5254 & 1.36   & 0.9706 & 0.76(0.13) & 1.20(0.12) & 1.15(0.12) & \nodata  \nl
Q1727+5311 & 1.81   & 1.1536 & 1.26(0.39) & 2.32(0.45) & 2.38(0.41) & \nodata \nl
\enddata
\end{deluxetable}
\end{center}

\clearpage

\begin{figure}
\plotone{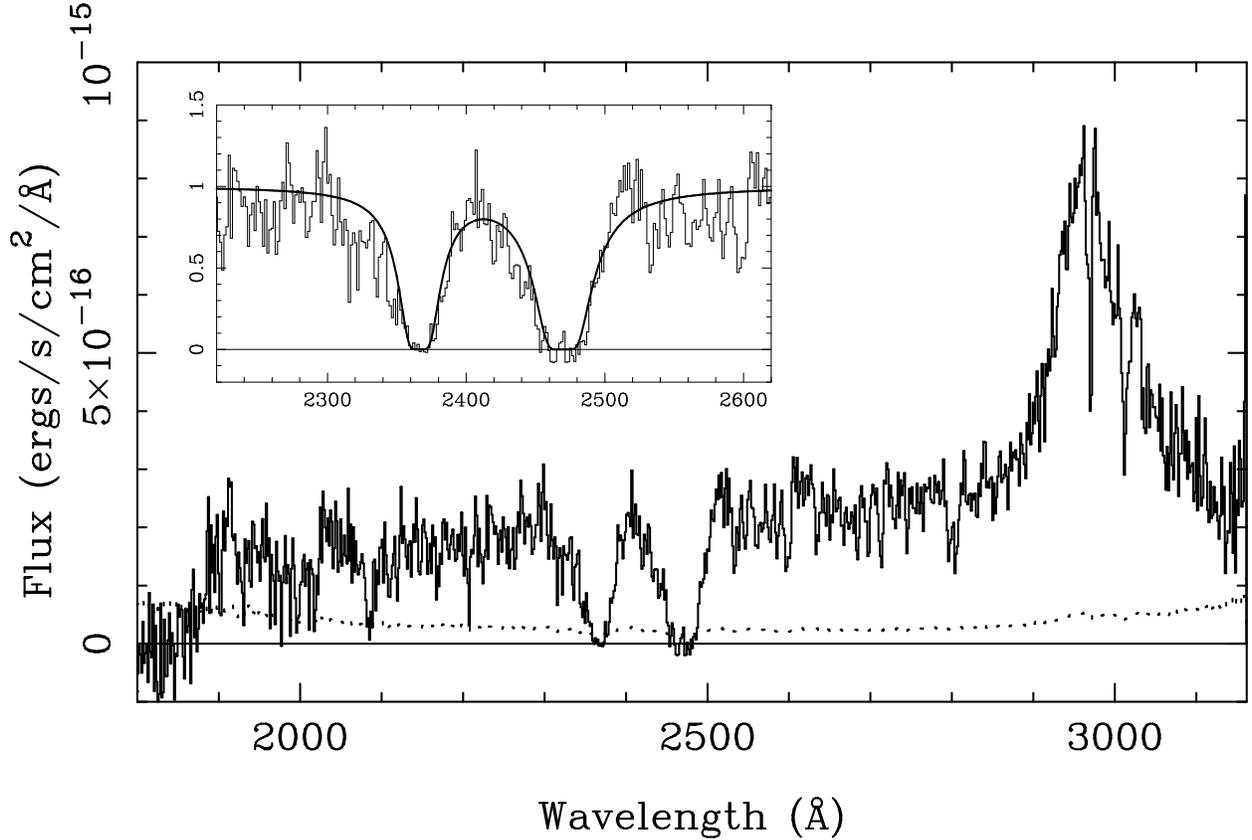}
\caption{HST UV spectrum of the SDSS $z_{em}=1.44$ quasar Q1727+5302
taken with the STIS NUV/MAMA G230L grating.  The quasar's Ly$\alpha$
emission line is at 2966\AA\ and the double-damped Ly$\alpha$
lines are at 2366\AA\ and 2470\AA. The dotted line is the $1\sigma$
error in flux.  The insert shows Voigt
profile fits overlaid on the two DLA absorption
lines at $z=0.9465$ and $z=1.0315$, with \ion{H}{1}
column densities of $N_{HI}=1.45\times10^{21}$ atoms cm$^{-2}$
and $N_{HI}=2.60\times10^{21}$ atoms cm$^{-2}$, respectively. 
The discrepancy between the fitted and observed spectra
in the long-wavelength wing of the higher redshift system
is probably due to weak Ly$\beta$/OVI emission from
the quasar which was not modeled in the fit.}
\end{figure}

\begin{figure}
\plotone{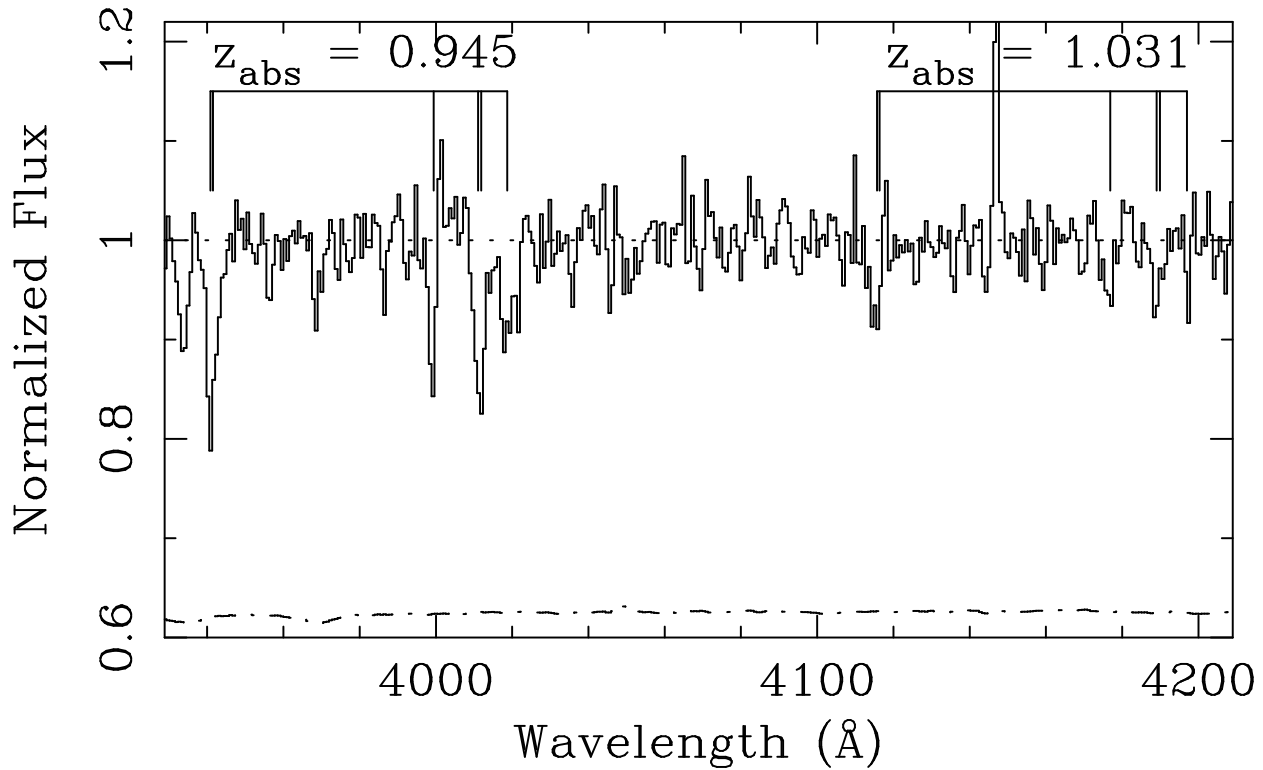}
\caption{MMT spectrum showing the \ion{Zn}{2} and \ion{Cr}{2} lines in 
both systems of the double-damped absorption. A discussion about the abundance 
determinations is given in \S2.1. Note that the scale shows only the top 
part of the normalized spectrum and 0.6 has been added to the error array 
(dash-dotted line). From left to right the lines labeled in each system 
are due to 
\ion{Zn}{2} $\lambda2026$ (blended with very weak \ion{Mg}{1} $\lambda2026$), 
\ion{Cr}{2} $\lambda2056$, a blend of \ion{Zn}{2} $\lambda2062$ and 
\ion{Cr}{2} $\lambda2062$, and \ion{Cr}{2} $\lambda2066$.}
\end{figure}

\begin{figure}
\plotone{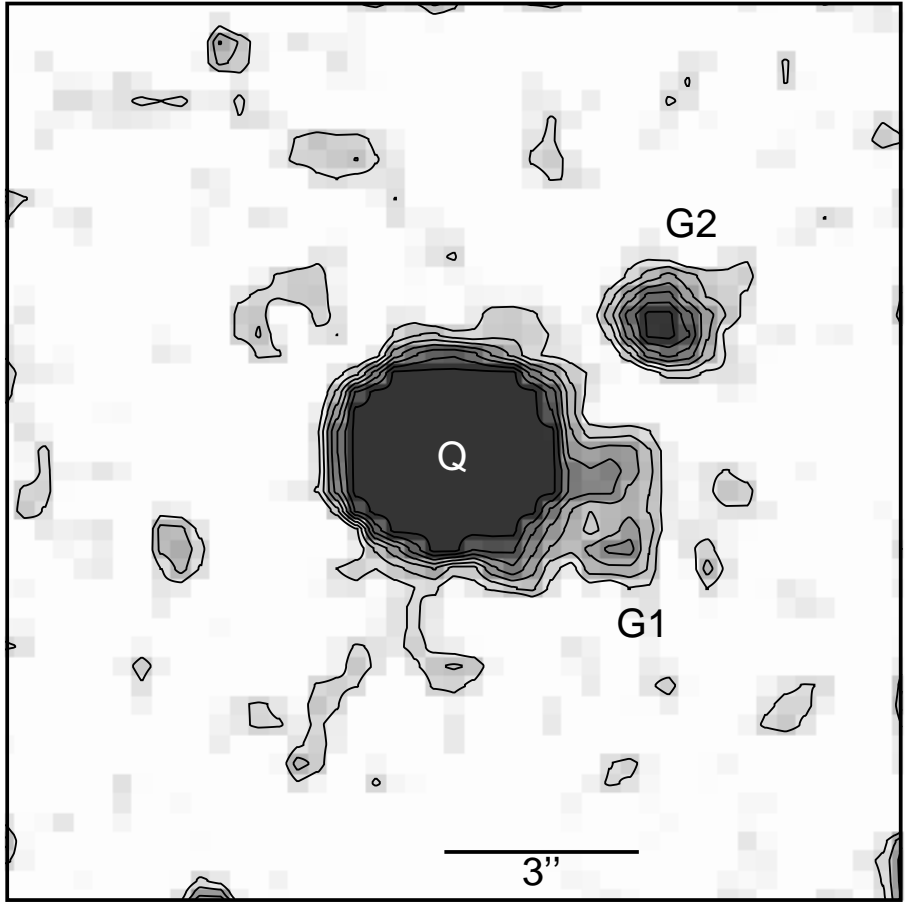}
\caption{Isophotal plot overlaid on an H-band image of the 
double-damped sightline taken in 1.1 arcsec seeing with the 3.0-m NASA IRTF 
telescope.
It has been smoothed to highlight
low surface brightness features, however, measurements were made
on the unsmoothed image. 
The image is 14\arcsec\ x 14\arcsec\ centered on the quasar (Q). 
G1 and G2 are the nearest resolved objects with impact parameters
of 2.7\arcsec\ and 4.0\arcsec. If at $z=1$, these correspond to proper 
transverse distances of 22$h_{70}^{-1}$ kpc and 32$h_{70}^{-1}$ kpc, 
respectively.}
\end{figure}

\begin{figure}
\rotate
\epsscale{0.9}
\plotone{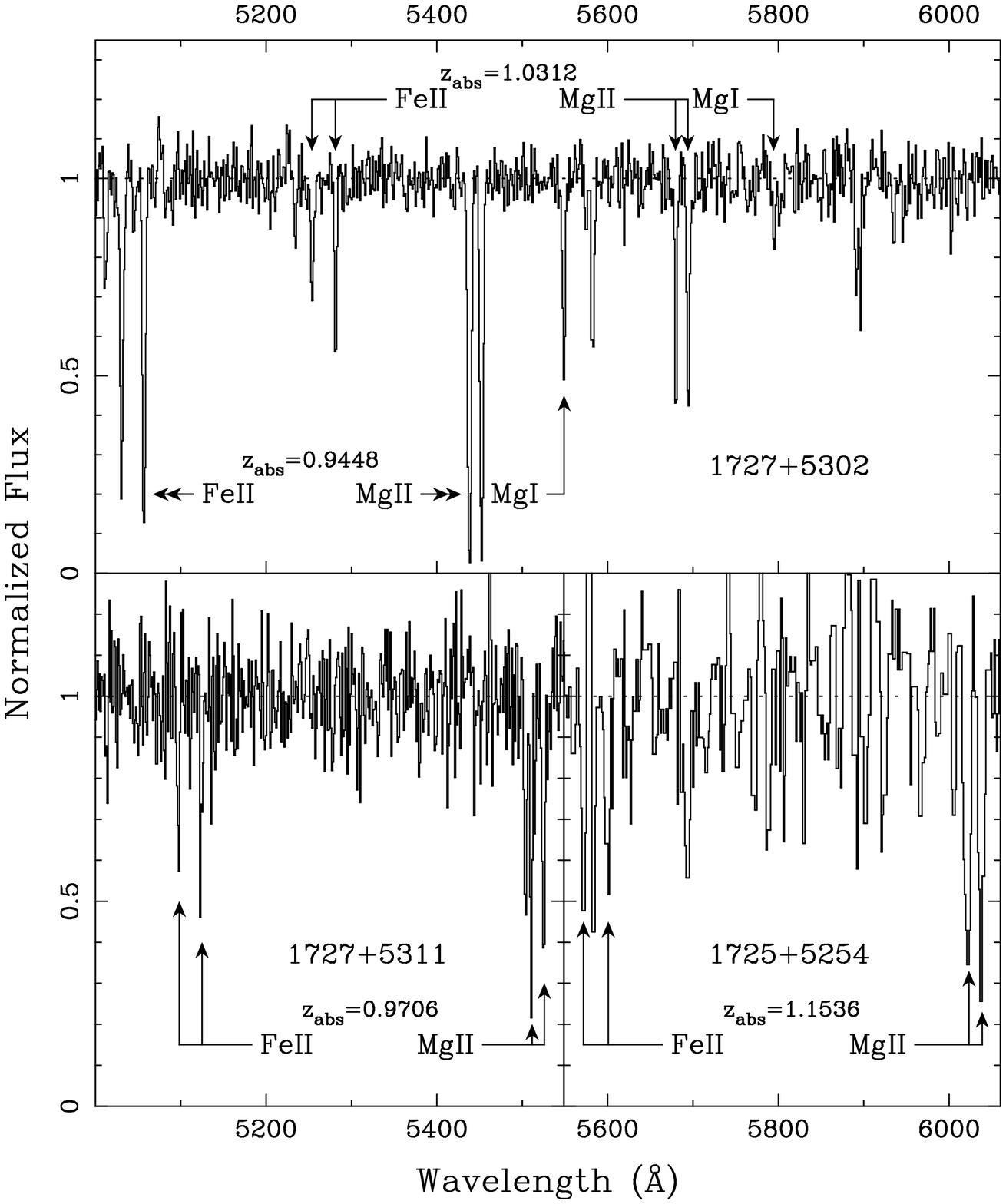}
\caption{Regions of the SDSS spectra for the three quasars 
which show strong \ion{Mg}{2}-\ion{Fe}{2} absorption with the continua
normalized. The top panel shows the two strong \ion{Mg}{2}-\ion{Fe}{2} systems
that are DLAs in Q1727+5302 at $z=0.9448$ and $z=1.0312$.
The bottom two panels show spectra of two SDSS quasars near 
the double-damped sightline. The bottom left panel shows 
the strong \ion{Mg}{2}-\ion{Fe}{2} system at $z=0.9706$ in Q1727+5311.
The bottom right panel shows the strong \ion{Mg}{2}-\ion{Fe}{2} system
at $z=1.1536$ in Q1725+5254.}
\end{figure}


\begin{thebibliography}{}

\bibitem[] {}
Adelberger, K., Steidel, C., Shapley, A., \& Pettini, M. 2003, ApJ, 584, 45

\bibitem[] {}
Bell, E. F., McIntosh, D. H., Katz, N., \&  Weinberg, M. D. 2003,
ApJS, 149, 289

\bibitem[] {}
Bouch\'e, N., \& Lowenthal, J. 2003, ApJ, 596, 810

\bibitem[] {} 
Eisenstein, D., Loeb, A., \& Turner, E. 1997, ApJ, 475, 421

\bibitem[] {}
Ellis, S., \& Jones, L. 2004, MNRAS, 348, 165

\bibitem[] {}
Faltenbacher, A., Gottlober, S., Kerscher, M., \& Muller, V. 2002, A\&A, 395, 1

\bibitem[] {}
Gawiser, E.,  et al. 2001, ApJ, 562, 628

\bibitem[] {}
Grevesse, N. \& Sauval, A. 1998,
Space Science Reviews, 85, 1/2, 161

\bibitem[] {}
Hamilton, T. S., Casertano, S., \& Turnshek, D. A. 2002, ApJ, 576, 61

\bibitem[] {}
Jahnke, K. \& Wisotzki, L. 2003, MNRAS, 346, 304

\bibitem[] {}
Miller, S. M., et al. 2004, MNRAS, submitted (astro-ph/0403065)

\bibitem[] {} 
Nestor, D., Rao, S., Turnshek, D., \& Vanden Berk, D.
2003, ApJ, 595, L5

\bibitem[] {}
Poggianti, B. M. 1997, A\&AS, 122, 399

\bibitem[] {} 
Prochaska, J. X., et al. 2003, ApJ, 595, L9

\bibitem[] {} 
Schneider, D., et al. 2002, AJ, 123, 567

\bibitem[] {} 
Rao, S., \& Turnshek, D. 2000, ApJS, 130, 1 (RT2000)

\bibitem[] {}
Rao, S., Nestor, D., Turnshek, D.,  Lane, W., Monier, E.,
\& Bergeron, J. 2003, ApJ, 595, 94

\bibitem[] {}
Rao, S., Turnshek, D., \& Nestor, D. 2004, in preparation (RTN2004)

\bibitem[] {}
Spergel, D., et al. 2003, ApJS, 148, 175

\bibitem[]{}
Palunas, P., Teplitz, H., Francis, P., Williger, G., Woodgate, B.
2004, ApJ, 602, 545

\end{thebibliography}
\end{document}